# Bioinformatics analysis of experimentally determined protein complexes in the yeast, *S. cerevisiae*

Zoltán Dezső[1], Zoltán N. Oltvai[2] and Albert-László Barabási[1]

[1]*Department of Physics, University of Notre Dame, Notre Dame, IN 46556, USA*

[2]*Department of Pathology, Northwestern University, Chicago, Illinois 60611, USA*

**Many important cellular functions are implemented by protein complexes that act as sophisticated molecular machines of varying size and temporal stability. Here we demonstrate quantitatively that protein complexes in the yeast, *Saccharomyces cerevisiae*, are comprised of a core in which subunits are highly coexpressed, display the same deletion phenotype (essential or non-essential) and share identical functional classification and cellular localization. This core is surrounded by a functionally mixed group of proteins, which likely represent short-lived- or spurious attachments. The results allow us to define the deletion phenotype and cellular task of most known complexes, and to identify with high confidence the biochemical role of hundreds of proteins with yet unassigned functionality.**

INTRODUCTION

Large-scale mass-spectrometric studies in *S. cerevisiae* provide a compendium of protein complexes (Alberts 1998, Hartwell 1999) that are considered to play a key role in carrying out yeast functionality (Gavin 2002, Ho 2002). While vastly informative, such libraries offer information only on the composition of a protein complex at a given time and developmental- or environmental condition. In addition, mass spectrometry is unable to distinguish those subunits that carry the key functional modules (i.e., the *core*) of the complex from those structural subunits that represent short-lived modulatory- or spurious associations (Von Mering 2002). Repeated individual purifications coupled with e.g., crystallographic- or cyro-electron microscopy characterization of each of these complexes could offer a more precise picture (Frank 2001, Abbott 2002), but such approaches on a large-scale are unavailable at present. Yet, extensive datasets on the essentiality, cellular localization and functional role of individual proteins, together with their corresponding gene expression, may allow us to develop an insight into the organization of protein complexes, and to provide a new perspective on the role of the various protein subunits.

RESULTS

We start by demonstrating that the cellular role and essentiality of a protein complex may largely be determined by a small group of protein subunits that display a high mRNA coexpression pattern, belong to the same functional class, and share the same deletion phenotype and cellular localization. For each $i$ and $j$ protein pair of an experimentally identified $N$ protein complex we calculated their corresponding mRNA coexpression coefficient (Eisen 1998), $\phi_{ij}$, that approximates the average coexpression coefficient of protein $i$ with all other subunits of the complex (Futcher 1999). We determined separately $C_i^D$ from global microarray data obtained on individual gene deletion mutants (Winzeler 1999, Hughes 2000), and $C_i^C$ from time kinetic data obtained on the yeast cell cycle (Cho 1998, Spellman 1998). The average correlation coefficient for each of the protein subunits of six large complexes (from Gavin *et al.*) is shown in the first columns of Figure 1. We find that a significant fraction of the protein subunits display a large, positive average mRNA coexpression coefficient with each other, indicating their potential functional relatedness to the other subunits within the complex. This result is in agreement with earlier findings of correlation between protein-protein interaction and transcriptional profiles (Grigoriev 2001, Ge 2001, Mrowka 2001, Jansen 2002, Kemmeren 2002). Some subunits, however, possess close to zero or even a negative correlation coefficient with the other subunits, indicating that they are not consistently coexpressed with the other subunits within the complex.

The internal correlations among the subunits of a protein complex are best revealed using a two dimensional representation, plotting for each protein $i$ the correlation coefficient $C_i^D$ on one axis and $C_i^C$ on the other. On such a plot, we color code each protein using essentiality information based on single gene deletions (column II in Fig.1), on the proteins' functional role (column III in Fig.1) and their known cellular localization (column IV in Fig.1), based on information compiled by the MIPS database (Mewes 2002). Such plots indicate the existence of two types of protein complexes, to which we refer to as *essential* (Fig.1a) and *non-essential* (Fig. 1b) complexes. For essential complexes we observe a mostly clear separation between the many essential and few non-essential protein subunits. For example in the three complexes shown in Fig. 1a, essential proteins aggregate in the high coexpression region of the mRNA coexpression phase space. A similar separation is observed for the non-essential complexes as well (Fig. 1a), where non-essential proteins aggregate in the high coexpression region. Finally, while most proteins belong to several functional classes, we find that for each complex displayed in Fig. 1 the vast majority of the highly coexpressed proteins share the same functional class and subcellular localization (Fig. 1, column III & IV).

To quantify the observed essentiality-, functional role- and cellular localization based separation we denote by $\overline{C}^D$ and $\overline{C}^C$ the average coexpression coefficient, obtained by averaging $C_i^D$ and $C_i^C$ over all proteins within a given complex, and by $\sigma^D$ or $\sigma^C$ the standard deviation around the average. We assume that all protein subunits $i$ for which $C_i^D > \overline{C}^D - \sigma^D$ and $C_i^C > \overline{C}^C - \sigma^C$ are part of the *core* of the protein complex. The protein subunits satisfying this condition are those depicted in the shaded areas in Fig. 1, allowing us to separate the core proteins from those that show only weak correlation with the other components of the complex. As Fig. 1 shows, we find that the core is characterized by a surprising degree of functional, essentiality and localization homogeneity: for example, of the forty proteins within the core of the complexes shown in Figure 1a thirty eight are essential. In addition, all core subunits share the same functional classification and cellular localization. Similarly, for the three complexes shown in Fig. 1b of the 49 core proteins only one is an essential protein; only four proteins with known functional role do not share the function of the majority; and all proteins share their cellular localization with the majority within the core. As the Supplementary Material demonstrates, where we list similar plots for 132 additional complexes, an essentiality-, function- and localization based homogeneity of the core is a generic property of most protein complexes.

The relatively unambiguous segregation of the essential and non-essential proteins within the complexes suggest that protein complexes may be categorized according to the deletion phenotype of the majority of their core subunits. Here we consider a specific complex essential if ≥60% of the core proteins with known deletion phenotype are essential, and non-essential if ≥60% of the core subunits are non-essential. We find that of the 383 complexes identified by Gavin *et al.* with three or more protein subunits (Gavin 2002), 174 are essential, 155 are non-essential, and only 54 do not show a clear classification based on the deletion phenotype of the core. Yet, a closer inspection indicates the majority of these 54 complexes are in fact non-essential. Indeed, most essential proteins found in the core of the ambiguous complexes participate in the core of other unambiguously essential complexes (see square symbols in Column II of Fig. 1b), indicating that their essentiality likely stems from their association with other essential complexes. When not considering these subunits we find that 35 of the 54 complexes with previously unclear classification are in fact non-essential. We also expect that the remaining 19 unclassified complexes could be also unambiguously classified as non-essential once a more complete list of all essential complexes becomes available. The Supplementary Material provide detailed predictions on the characteristics of all complexes identified by Gavin *et al.* (Gavin 2002), Ho *et al.* (Ho 2002), and those collected in the MIPS database (Mewes 2002). In addition, when we computationally simulate subunit compositions identical in numbers with those identified experimentally by Gavin *et al.* (Gavin 2002), but whose composition is selected randomly from the yeast proteome, we derive only 9 essential complexes (Fig. 2a), indicating that the experimentally identified complex ensemble is highly non-random and is biased towards essential complexes. As a specific example in Supplementary Material we show a negative control set of Fig. 1, with randomly selected proteins, indicating the absence of functional and essentiality based separation of the core and halo proteins.

The results also indicate a relatively uneven distribution of the essential complexes in different functional categories and localization classes. Indeed, we find that the majority of protein complexes are responsible for subcellular localization and transcription (Fig. 2c), and are located in the nucleus and cytoplasm (Fig. 2d). This is consistent with the known bias of mass-spectrometry approaches towards nuclear proteins (Von Mering 2002). Interestingly, in the nucleus the essential complexes outnumber the non-essential complexes, a bias that is inverted in the cytoplasm-associated complexes. Finally, we find a weak, but positive correlation between the size of the complex and its essentiality: the larger the complex, the more likely that its core is essential (Fig. 2b). For example, only ~45% of the complexes identified by Gavin *et al.* (Gavin 2002) with 10 or less proteins are essential. This fraction increases to 100% for complexes with more than 40 subunits.

DISCUSSION

Many biological functions are carried out by the integrated activity of highly interacting cellular components, referred to as functional modules. Here we investigated the properties of one type of such modules; the protein complexes found in *S. cerevisiae*. Our results suggest that many of the identified protein complexes possess an invariant core, in which the biochemical role of each protein subunit is irreplaceable, and is seamlessly integrated into a higher-level function of the whole complex. In turn, the deletion phenotype of each core protein is determined by the role of the complex in the organism. If the given complex is essential for cell growth, the deletion of *any* core protein disrupts the complex's functional integrity, and subsequently renders the cell unviable (Fig. 2e). If however, the cell is able to tolerate the loss of a complex's function, none of its specific core subunits are essential (Fig. 2f). The core is generally surrounded by several 'halo' proteins that typically do not share a common deletion phenotype, functional classification or cellular localization with the core subunits (Fig. 2e,f). This indicates that they likely represent temporal attachments, some acting as modifiers of the complex' function, while others are functionally unrelated proteins that spuriously attach to the surface of the core proteins (Von Mering 2002).

Our ability to identify the core, together with the observed essentiality, functional and localization based homogeneity of the core, allows a more precise identification of those subunits for which a possible cellular function can be inferred (Gavin 2002, Ho 2002) (See Supplementary Material). Indeed, participation in a specific complex can be considered as source of functional classification. Our results indicate, however, that such functional assignment can be made with high confidence only for the core proteins. To turn our findings into a predictive tool, we identified all proteins that belong to the core of a large complex, and have either an unknown functional classification or one whose current functional annotation differs from the majority of the other core proteins in the complex. This identification allowed as to assign functional prediction to 869 core proteins listed in Table II, IV and VI in the Supplementary Material.

The segregation of protein complexes into essential and non-essential ones offers a new perspective on the organizational level at which a protein's deletion phenotype is determined. Based on data, it is evident that to a high degree a protein's phenotypic essentiality is determined by the role it plays in ensuring the integrity of vital molecular complexes, thus elevating essentiality from the property of an individual protein (Jeong 2001) to a characteristic of the protein complex. In agreement with this proposition, we find that almost 47% (508) of all known essential yeast proteins (1085) are part of the core of complexes identified by Gavin *et al*. (Gavin 2002), despite the fact that the total number of proteins in these complexes represent only ~20% (1363) of all yeast proteins (6316). Presumably, a complete list of protein complexes could associate an even larger fraction of essential proteins with such essential complexes. This internal organization is consistent with the notion of stable or unstable protein complexes (Jansen 2002), and the dynamical coexpression of selected open reading frames (Alter 2000, Holter 2000, Ge 2001). Understanding, the dynamics of the complex genetic networks (Hasty 2001, Solé 2002), potentially responsible for synchronizing the expression of the core subunits, is now a prime challenge.

METHODS

**Protein Complexes:** We used the complete list of protein complexes identified by Gavin *et al.* (Table S1 in Gavin *et al.*), by Ho *et al.* (Table S1 in Ho *et al.*) and the MIPS database (http://mips.gsf.de). We focus on complexes of three or more proteins, which limits us to 384, 585 and 144 complexes for the three databases, respectively. Each complex is identified based on the order number given in the original databases, and reproduced in the Supplementary Material. The deletion phenotype data for individual ORFs was downloaded from www-deletion.stanford.edu (version: June, 2002), while the functional classification and cellular localization of the individual proteins was obtained from the MIPS database (as of June, 2002).

**Coexpression Patterns:** The global mRNA expression data for yeast was downloaded from http://www.rii.com/tech/pubs/cell_hughes.htm (Hughes 2000) limiting our analysis to the genomic expression program of 287 single gene deletion mutant *S. cerevisiae* strains grown under identical cell culture conditions as wild-type yeast cells. A similar analysis was performed on the cell cycle datasets (Cho 1998, Spellman 1998). For each protein belonging to a given complex we determined $\phi_{ij}$, following Eisen *et al.* (Eisen 1998). The obtained $\phi^C_{ij}$ encodes the coexpression matrix based on the cell cycle data, and $\phi^D_{ij}$ based on the deletion datasets. The coefficient $C_i^{C,D} = (\sum_j \phi_{ij}^{C,D})/N$, where $N$ denotes the number of proteins in the studied complex, quantifies the average coexpression with the rest of the proteins within the complex, was determined for each protein subunits of all known complexes. The typical correlation value range between $-0.5$ and $0.5$. Note that for pairwise protein-protein interactions occasionally higher correlation coefficients are observed (Grigoriev 2001, Ge 2001, Mrowka 2001, Jansen 2002,

Kemmern 2002), a difference rooted in the fact that $C_i$ reflects the average correlation with all other complex subunits, some proteins contributing with small or negative values.

**Functional prediction**: We assign to each complex the functional role (cellular localization) shared by the majority of the core proteins. The confidence level of each prediction is based on the percentage of the core proteins known to belong to the selected functional class. Next, we identify all core proteins that either do not have a known functional classification, or their functional classification does not agree with the predicted functional role of the protein complex core in which they participate. For these proteins, based on the association with the core, we assign the functional role/cellular localization as predicted by the complex's role. Halo proteins are not included in this prediction process, as they do not display the functional and phenotype homogeneity seen in the core.

**Acknowledgements:** Research at Northwestern University and the University of Notre Dame was supported by the Department of Energy and by the National Institute of Health.

Correspondence and requests for materials should be addressed to A.-L. B. (alb@nd.edu) or Z.N.O. (zno008@northwestern.edu)

Figure 1

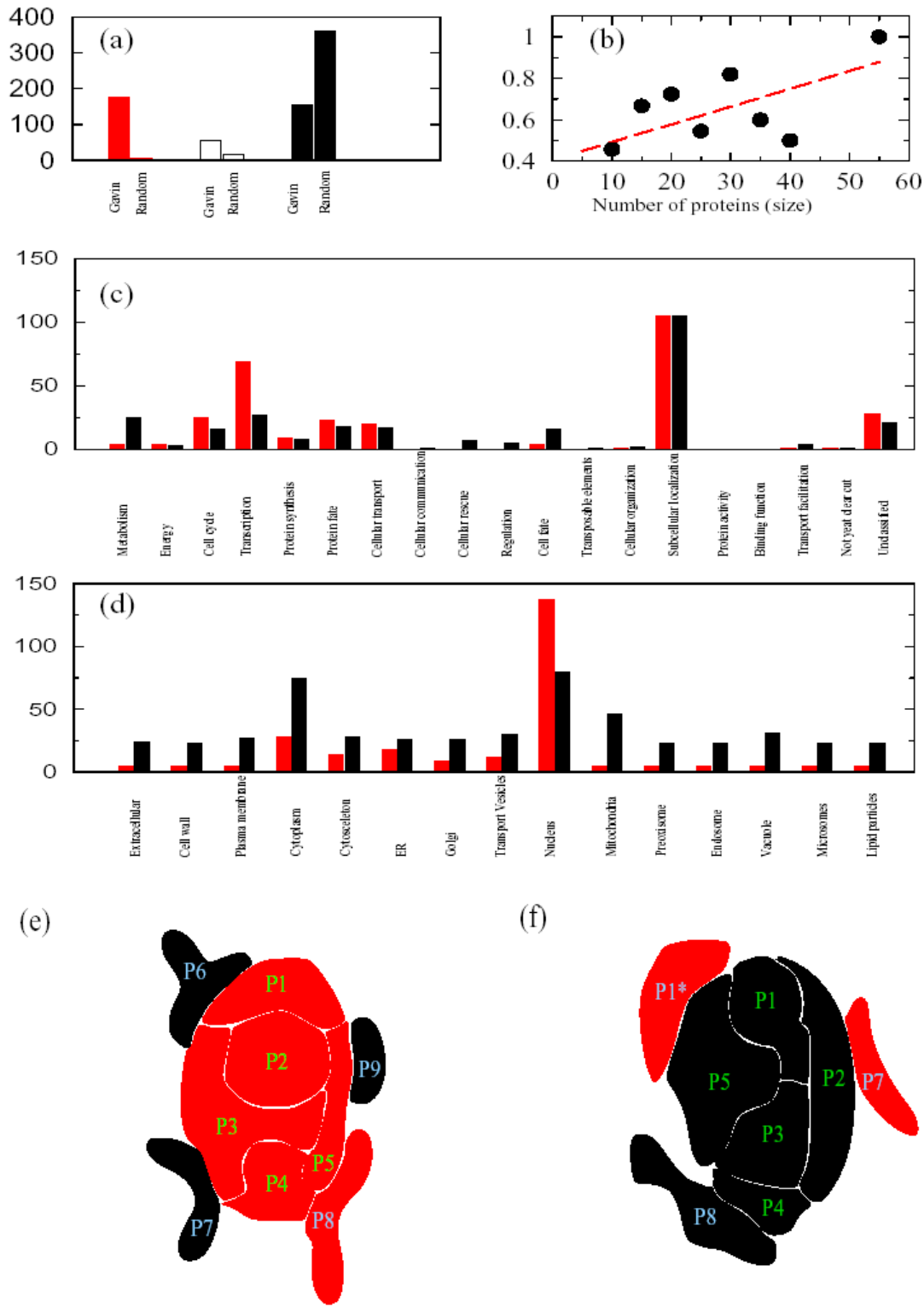

Figure 2

Figure Legends

**Figure 1    Characterizing three essential (a) and non-essential (b) complexes.**

**(a)** *Column I*: mRNA coexpression patterns for three large complexes identified in Gavin *et al.*. For each protein subunit (identified at the bottom of each panel) we show the average correlation coefficient for their corresponding relative mRNA expression level with all other subunits based on the microarray data obtained on gene deletion mutants (Winzeler 1999, Hughes 2000) ($C^D$, top plot), and cell-cycle measurements (Cho 1998, Spellman 1998) ($C^C$, bottom plot). We denote by red (black) the known essential (non-essential) proteins. *Column II*: Cross correlation plot obtained by plotting for each protein *i* within the three selected complexes the cell-cycle correlation coefficient $C_i^C$ on the horizontal axis, and the gene deletion correlation coefficient $C_i^D$ on the vertical axis. Each symbol corresponds to a single gene product (protein), the color reflecting its known deletion phenotype (red: essential; black: non-essential). The shaded area separates the highly coexpressed core proteins, the boundaries of the area being given by $C_i^C = \overline{C}^C - \sigma^C$ and $C_i^D = \overline{C}^D - \sigma^D$. *Column III*: The same coexpression plot as in Column II, but the symbols are color-coded based on the functional classification of the corresponding proteins. The green symbols denote gene products that belong to the majority regarding their known functional role (Complex 365 and 360: green proteins simultaneously belong to protein fate and subcellular localization; Complex 363: transcription) unfilled symbols denote proteins with unknown functional role; and the blue symbols denote those subunits that do not share the functional classification with the majority. *Column IV:*  Coexpression plot with proteins colored based on their known cellular localization. Green symbols denote those with the same subcellular localization, which is nucleus for all three complexes. Blue symbols denote proteins whose localization differs from the majority and unfilled symbols represent those with unknown cellular localization. In column II-IV we used a two dimensional representation to demonstrate that the essentiality, functional classification and cellular localization based separation is simultaneously present by using two widely different transcriptional datasets. A control plot with only the Cho *et al.* cell cycle data is shown in the Supplementary Material.

**(b)** The same as (a), but for three complexes with predominantly non-essential subunits. In *Column II* we used red squares to denote those essential proteins that are part of the core of other essential complexes. In *Column III* the green symbols represent protein participating in synthesis. In *Column IV* the green symbols denote proteins localized in the mitochondria for all three complexes.

**Figure 2   Characterization of the protein complex ensemble and schematic illustration of the internal organization of protein complexes:**

**(a)** The number of complexes in the Gavin *et al.* dataset (Gavin 2002) that are found to be essential (red), non-essential (black) and of unknown (white) deletion phenotype. Next to each column we show the number of corresponding complexes if the proteins were randomly distributed in the various complexes, indicating the highly non-random character of the complex composition and essentiality. For this each protein subunit of the known Gavin *et al.* complexes are replaced with proteins randomly selected from the yeast proteome. **(b)** The size dependence of essentiality in protein complexes. The plot shows the fraction of essential subunits within the complex (vertical axis) as in function of the number of protein subunits in the complexes (horizontal axis). **(c)** The predicted functional classification of the complexes identified by Gavin *et al.* (Gavin 2002), showing separately the number of essential and non-essential complexes found in each functional class. **(d)** The predicted cellular localization of the identified essential and non-essential complexes. A full list of predictions for each complex is shown in the Supplementary Material. **(e)** We find that approximately 43% of the protein complexes possess a core comprised of highly coexpressed proteins, that are all essential and belong to the same functional class, suggesting that they represent the functional building blocks of the complex. Such core is shown schematically as tightly locked P1-P5 proteins. Mass spectroscopic methods inevitably identify other proteins as well with those complexes. Yet, we find that these halo proteins (P6-P9) show a small coexpression pattern with the core, and are both phenotypically and functionally mixed, indicating that they likely represent proteins that display only temporal- or spurious attachment to the complex. **(f)** Approximately 46% of complexes have a core composed of predominantly non-essential proteins (P1-P5), surrounded again by a halo of proteins with mixed essentiality and functional classification (P6-P8). These complexes likely are not essential for cell growth, therefore all core proteins are uniformly non-essential. The few essential proteins found predominantly in the halo of such non-essential complexes often simultaneously take part in the core of other essential complexes, explaining the origin of their essentiality. For example, the P1 protein, which is part of the core of the essential complex shown in (e), could also attach to the surface of the non-essential complex shown in *b*. Therefore, the essentiality of P1* is derived not from its role in complex (f), but from its role in the essential complex (e).